\documentclass{aa501}
\usepackage{graphics,amssymb,natbib,rotating,psfrag,amsmath,epsfig,array}

\title{Timing diagrams and correlations in gamma-ray bursts signal jets
from accretion into black holes}
\titlerunning{Timing diagrams in GRBs}

\author{S.\,McBreen \and
        B.\,McBreen \and
        F.\,Quilligan \and
        L.\,Hanlon}

 \institute{Department of Experimental Physics, University
College Dublin, Dublin 4, Ireland}

\offprints{smcbreen@bermuda.ucd.ie}
\date{Received 22 October 2001 / Accepted 10 December 2001}

\abstract{The temporal properties of a sample of 498 bright
gamma-ray bursts (GRBs) with durations between 0.05 s and 674 s
were analysed.  The large range in duration (T$_{90}$) is
accompanied by a similarly large range in the median values of
the pulse timing properties including rise time, fall time, FWHM
and separation between the pulses.  Four timing diagrams relating
these pulse properties to T$_{90}$ are presented and show the
power law relationships between the median values of the 4 pulse
timing properties and T$_{90}$, but also that the power laws
depend in a consistent manner on the number of pulses per GRB.
The timing diagrams are caused by the correlated properties of
the pulses in the burst and can be explained by a combination of
factors including the Doppler boost factor $\Gamma$, a viewing
effect caused by a jet and different progenitors.  GRBs with
similar values of T$_{90}$ have a wide range in the number of
pulses. GRBs with the large number of short and spectrally hard
pulses may occur either from a homogeneous jet with a higher
average value of $\Gamma$ or close to the axis of an
inhomogeneous jet with higher values of $\Gamma$ near the
rotation axis. The less luminous GRBs with fewer pulses may
originate further from the axis of the inhomogeneous jet. The
pulses in GRBs have six distinctive statistical properties
including correlations between time intervals, correlations
between pulse amplitudes, an anticorrelation between pulse
amplitudes and time intervals, and a link to intermittency in GRS
1915+105. The timing diagrams and correlated pulses suggest that
GRBs are powered by accretion processes signalling jets from the
formation of black holes.
 }

\begin{document}
 \maketitle

\section{Introduction}
The cosmological origin of gamma-ray bursts (GRBs) requires an
extraordinary amount of energy to be emitted in gamma-rays
\citep{piran:1999}. The source of this energy is thought to be a
cataclysmic event involving mergers of compact objects such as
neutron stars or a neutron star and black hole
\citep{pacz:1991,ruffjan:1999} or the formation of a black hole
in massive stars \citep{macfad:1999,popham:1999}. GRB light
curves are complex and irregular \citep{fishman:1995} and a range
of techniques have been developed to elucidate their structure
\citep{mhlm:1994,nnb:1996,lee:2000,belss:2000,quilligan:2002}. In
this paper we present the results of the analysis of pulses in a
large sample including long and short GRBs. The main results are
presented in section 3 including four timing diagrams and our
analysis of pulse properties. These results along with the unique
properties of the pulses in GRBs are interpreted in section 4 and
used to provide additional insights to the emission process and
the central engine that support jets from accretion into newly
formed black holes.

\begin{figure*}
\resizebox{\textwidth}{!}{\rotatebox{-90}{\includegraphics{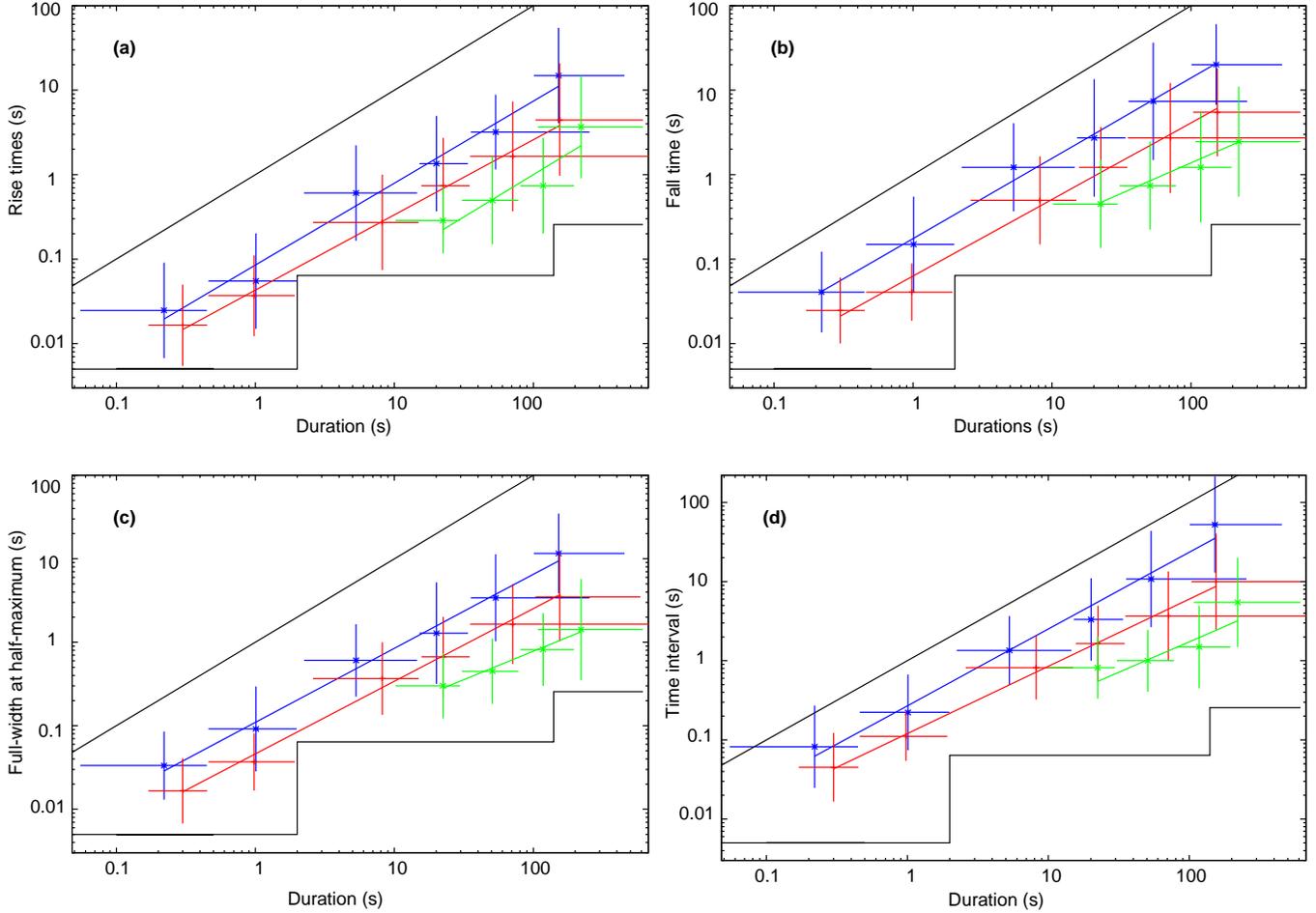}}}
\label{fig:grbs_pulses} \vspace{.5em} \caption{The timing
diagrams in GRBs. The median values obtained from the lognormal
distributions for the pulse timing parameters (a) t$_{\rm r}$, (b)
t$_{\rm f}$, (c) FWHM and (d) $\Delta$T are plotted versus
duration T$_{90}$ for GRBs in three categories i.e. 1 $\leq$ N
$\leq$ 3 (blue), 4 $\leq$ N $\leq$ 12 (red) and N $>$ 12 (green).
The crosses signify the range covered in T90 and \(e^{\mu \pm
\sigma}\) for the lognormal distribution which includes 16\% to
84\% of the pulse values for that T$_{90}$ bin.  The upper
diagonal line is the limit where the pulse parameter is equal to
T$_{90}$ and the lower lines by the limited time resolution i.e.
5 ms, 64 ms and 256 ms. In some GRBs with few pulses separated by
a long $\Delta$T the value of T$_{90}$ which goes from 5\% to
95\% of the counts may not include a small peak before or after
the $\Delta$T yielding a value of $\Delta$T $>$ T$_{90}$. This
effect is seen in (d).  Small corrections were applied to the
pulse timing locations of the crosses  that took into account the
truncation of the distributions by the limited time resolution.
The lines are the best fits to the median values and the power
law indices are listed in Table 1. The timing values were
obtained for pulses that were isolated from neighbouring pulses
at $\geq$ 50\% whereas all pulses above 5 $\sigma$ were used for
$\Delta$T.}
\end{figure*}

\section{Pulse analysis in GRBs}

The BATSE experiment has provided the most uniform collection of
GRBs available.  A large sample of the brightest GRBs with data
combined from the four energy channels was analysed using either
wavelets or median filters.  A detailed account of this process
has been described elsewhere \citep{quilligan:2002,sheila:2001}.
An automated pulse selection algorithm was used to detect pulses
in the denoised signals.  The samples included the brightest 319
GRBs with T$_{90}
>$ 2 s analysed at 64 ms resolution and 100 GRBs with T$_{90} <$
2 s analysed at 5 ms resolution. To extend the sample to include
GRBs with T$_{90}$ well beyond 100 s, a further 79 bright GRBs
were included and analysed at 256 ms resolution.

Pulses that were isolated from neighbouring pulses at $\geq$50\%
level were selected and the rise times $t_{\rm r}$, fall times $t_{\rm f}$
and FWHM were measured.  The time intervals between pulses
$\Delta$T, above the 5$\sigma$ threshold, were determined.  The
individual distributions of t$_{\rm r}$, t$_{\rm f}$, FWHM and $\Delta$T
have a wide range and are consistent with lognormal
distributions.  The bursts were split into T$_{90}$
categories with comparable number of pulses and typical lower boundaries
 0, 0.5, 2, 15, 30 and greater than 100 s for the GRBs at 256 ms resolution.
 The latter GRBs were binned separately despite the overlap with
T$_{90}$ because they are from a dimmer sample that was included
to extend the range in duration to ultra-long GRBs. The GRBs were
further split into three categories based on the number of pulses
N $\geq$ 5$\sigma$ i.e. 1 $\leq$ N $\leq$ 3, 4 $\leq$ N $\leq$ 12
and N $\geq$ 12. The median $\mu$ and standard deviation $\sigma$
of each lognormal distribution was determined
\citep{quilligan:2002}.

\section{Results}

\begin{table}[b]
 \caption{Indices of the power law fits to the timing data.}
\setlength{\tabcolsep}{0.3cm}
\begin{tabular}[b]{|l|c|c|c|} \hline
%\vspace{0.1cm}
Property   & 1 $\leq$ N $\leq$ 3 & 4 $\leq$ N
$\leq$ 12 & N $>$ 12 \\
\hline
Rise Time       & 0.98 $\pm0.07$        & 0.89 $\pm0.03$        & 1.02 $\pm0.3$ \\
Fall Time       & 0.95 $\pm0.04$        & 0.91 $\pm0.05$        & 0.72 $\pm0.1$ \\
FWHM            & 0.89 $\pm0.04$        & 0.87 $\pm0.03$        & 0.68 $\pm0.06$ \\
Time Interval   & 0.97 $\pm0.06$        & 0.85 $\pm0.03$        & 0.77 $\pm0.27$  \\ \hline
\end{tabular}
\end{table}

The diagrams of t$_{\rm r}$, t$_{\rm f}$, FWHM and $\Delta$T
versus T$_{90}$ are presented in Fig. 1 for three categories
based on the number of pulses in the GRBs.  The crosses in Fig. 1
are not error bars but show the range in T$_{90}$ over which  the
pulses in the bursts were combined and the range of the lognormal
distribution that includes 67\% of the values for the relevant
pulse property. The indices of the best fit power-laws to the
median values in Fig. 1 are listed in Table 1 and all have values
close to unity. The values of t$_{\rm r}$, t$_{\rm f}$, FWHM and
$\Delta$T are well correlated with each other
\citep{quilligan:2002} which explains the similarity of the
diagrams when plotted versus T$_{90}$. All four figures are
presented to emphasise this result. For the two categories with 1
$\leq$ N $\leq$ 3 and 4 $\leq$ N $\leq$ 12 the power-laws between
the median values of t$_{\rm r}$, t$_{\rm f}$, FWHM, $\Delta$T
and T$_{90}$ extend over a range of more than 10$^{3}$ and
includes GRBs from both sub-classes.  The GRBs with N $>$ 12
include only those with T$_{90} >$ 2 s because N = 9 is the
largest value for GRBs with T$_{90} <$ 2 s.  The data are
displaced from the two catagories of GRBs with fewer pulses and
display flattening for T$_{90} <$ 50 s which may be caused by a
selection effect because of pulse amalgamation by the 64 ms
resolution of the data.

The properties of the isolated pulses in GRBs with T$_{90} >$ 2 s,
at 64 ms resolution, were compared with the preceding
($\Delta$T$^{-}$) and subsequent ($\Delta$T$^{+}$) values of
$\Delta$T.  The pulse amplitudes (PA) are more anticorrelated with
$\Delta$T$^{-}$ than $\Delta$T$^{+}$ with Spearman rank order
correlation coefficients of -0.39 and -0.33.  The peak to peak
values of $\Delta$T were then corrected for the strong
correlation between $\Delta$T and the pulse rise and fall times
\citep{quilligan:2002}. The amended values of the correlation
coefficients are -0.27 and -0.16 with probabilities that the
anti-correlations arose by chance of 10$^{-15}$ and 10$^{-6}$.
This difference did not depend on the isolation level of the
pulses which was varied from 30\% to 80\%.  The FWHM of the
pulses were also compared with $\Delta$T$^{-}$ and
$\Delta$T$^{+}$ and found to be strongly correlated.  The
correlations were marginally stronger for $\Delta$T$^{-}$ than
$\Delta$T$^{+}$.

\section{Discussion}
The data in Fig. 1 shows that the median values of t$_{\rm r}$,
t$_{\rm f}$, FWHM and $\Delta$T depend strongly on T$_{90}$ and
also on the number of pulses.  GRBs with the same durations can
have a wide variation in the number of pulses and the pulse
properties. GRBs with a small number of pulses generally have
slow pulses that are further apart while bursts with larger
number of pulses have faster and spectrally harder pulses that
are closer together \citep{norris:2001}. In both cases the same
pattern is observed except that the median values of t$_{\rm r}$,
t$_{\rm f}$, FWHM and $\Delta$T scale with T$_{90}$. Bursts with
only one pulse and correlated pulse properties would lie on a
line of slope unity, while bursts with many pulses lie on
parallel lines if the pulse properties and $\Delta$T are
correlated. The correlated nature of the pulse parameters and
$\Delta$T provide additional information on the central engine
and the emission process.

\subsection{Number of pulses, pulse properties and jets}

The separation of GRBs based on N (Fig. 1) suggests a kinematic
origin because the median values of the pulse timing parameters
and $\Delta$T scale in the same way and by about the same amount.
In the internal shock model, an effect of this type may occur in
homogeneous jet models where the degree of collimation and range
in $\Gamma$ depends on the mass at the explosion
\citep{kobay:2001} or in an inhomogeneous jet model by a viewing
effect caused by collimated emission with higher average values of
$\Gamma$ close to the rotation axis, where there is lower baryon
pollution \citep{reemes:1994,rees:1999,sal:2000}. The
transparency radius of the fireball r$_{t}$ varies as
$\Gamma^{-\frac{1}{2}}$.  The profiles of pulses in bursts from
off-axis will be slowed by (1) the reduced value of $\Gamma$ and
(2) the longer time for faster shocks to catch slower ones before
reaching r$_{t}$. There will be fewer shocks to collide outside
r$_{t}$ and hence generate lower luminosity bursts with slower
and softer pulses because the additional shock amalgamation
produces a narrower range in $\Gamma$. Indeed this process may be
a key factor in controlling the spread in $\Gamma$ in GRBs.

The luminosity of BATSE bursts is not determined but several
factors imply that GRBs with faster and harder pulses are more
luminous. The luminosity-variability and luminosity-lag
correlations for GRBs with known z infer that the more variable
and spectrally harder bursts are more luminous
\citep{nmb:2000,sal:2000,feniram:2001,schaefer:2001,ioka:2001,schmidt:2001}.
These correlations can be explained in an inhomogeneous jet model
where the more luminous and variable bursts with higher N occur
close to the axis and the slower, softer and less luminous bursts
from further off-axis \citep{quilligan:2002}. In addition, there
is the strong case for jets in GRBs based on the impressive range
of afterglow studies \citep{frail:2001}. The unexpected effect is
that the jet has left its imprint on the timing profile of the
burst.

\subsection{Variation of pulse properties and $\Delta$T with T$_{90}$}

The power laws in Fig. 1 connect the pulse data in the two
sub-classes for 1 $\leq$ N $\leq$ 3 and 3 $\leq$ N $\leq$ 12. The
median values of the pulse timing parameters and $\Delta$T scale
with T$_{90}$ by about a factor of 10$^{3}$. This result may
indicate a direct connection between the two sub-classes of
bursts or that they coincidentally lie on the same power laws in
the absence of sufficient data to independently determine the
slopes for the short bursts.  The popular progenitors of GRBs
range from mergers of compact objects such as neutron stars and
black holes to collapsars and hypernovae in massive stars. A neat
feature of most progenitor models is that they provide a generic
scenario based on the formation of a black hole with a massive
debris torus that is rapidly accreted and energises the jet by
neutrino transport and magnetohydrodynamic processes.

The variation of the pulse parameters with T$_{90}$ maybe caused
by the same emission mechanisms and progenitors.  If GRBs have
the same progenitors, the short GRBs must have very high values of
$\Gamma$ to produce the short durations \citep{piran:1999}. The
high value of $\Gamma$ is close to the upper bound allowed by the
internal shock model \citep{sapi:1997,lazzati:1999} but well
below what can in principle be produced in a very low baryon
environment \citep{meszar:1997}.  An alternative interpretation
is that there are two types of progenitors with pulses that
coincidentally lie on the same power laws. The progenitors that
arise from massive stars may not be capable of producing short
GRBs.  The short bursts may come from mergers of neutron stars or
neutron stars and black holes \citep{ruffjan:1999} where the time
structure in the bursts reflects the interaction of thin
relativistic shocks with $\Gamma \sim$ 100 and duration of $\sim$
50 ms.  In this two progenitor scenario the high values of
$\Gamma$ can be reduced by about an order of magnitude. In
addition the central engine and the environment before the
photosphere may combine to smooth the shocks in a way that is a
function of T$_{90}$.

\subsection{Correlations between pulses}

There are a number of new results on pulses, from short and long
GRBs, that constrain the emission process.  These include (1) the
distributions of values of t$_{\rm r}$, t$_{\rm f}$, FWHM, pulse
amplitude (PA) and pulse area of isolated pulses are not random
but have lognormal distributions
\citep{quilligan:2002,sheila:2001,np:2001}, (2) the values of
$\Delta$T are lognormally distributed with a power-law excess of
long time intervals or a Pareto-Levy tail for GRBs with T$_{90}
>$ 2 s \citep{quilligan:2002,sheila:2001,np:2001}, (3) there is an
anticorrelation between the values of PA and FWHM
\citep{ramfen:2000,quilligan:2002}, (4) there is a positive
correlation between the values of the PA that extend over many
pulses \citep{quilligan:2002,sheila:2001}, (5) there is also a
positive correlation between the values of $\Delta$T
\citep{quilligan:2002,sheila:2001}, (6) the PA of isolated pulses
in GRBs with T$_{90}
>$ 2 s and 64 ms resolution were compared with $\Delta$T$^{-}$ and
$\Delta$T$^{+}$ because of a strong correlation between the pulse
width and $\Delta$T$^{-}$ in the galactic superluminal source GRS
1915+105 where outbursts are powered by accretion into a black
hole \citep{belloni:1997} and a correlation between long quiescent
periods in GRBs and subsequent intervals of emission
\citep{rm:2001}. The PAs were found to be anticorrelated with
$\Delta$T$^{-}$ and $\Delta$T$^{+}$ with a slightly larger
anticorrelation for $\Delta$T$^{-}$.  Any additional correlations
between the pulse properties and $\Delta$T$^{-}$ would be
expected to be much weaker than in GRS 1915+105 because a similar
intermittent pattern in the outbursts from the central engine
would have to persist after shock interactions generated the
outbursts.  The combination of the six results make the pulses in
GRBs quite unique.  In addition long pulses are asymmetric and
reach their maximum earlier in higher energy bands while shorter
pulses tend to be more symmetric with negligible time lags
between energy channels \citep{nnb:1996}.

The three results discussed above seem to favour an internal
shock model powered by a hyper-accretion process into a newly
formed black hole \citep{ruffjan:1999,popham:1999,zhang:2001}.
The accretion process has been modelled and is sensitive to the
rate at which material piles up around the black hole and the
accretion time scale of the particles. The accretion rate might
provide the overall control on the process that eventually
generates the correlated outbursts from the jet. The pulse
amplitudes and time intervals can be coupled because higher
accretion rates cause larger outbursts that are closer in time,
while lower accretion rates produce smaller and slower events that
are further apart. It appears that variation in the rate of
accretion, the thickness of the relativistic shocks and the
viewing angle of the jet may be key factors in accounting for the
observed durations and pulse properties in GRBs. There are many
uncertain factors that influence the process. These include the
viscosity of the particles, the mass and angular momentum of the
disk, the mass and spin of the black hole and the energy
extraction and collimation process.
\bibliography{katmonic,Dj223}
\bibliographystyle{apj}
\end{document}